\documentclass[pra,aps,amsmath,twocolumn,showpacs,floatfix]{revtex4-1}
\usepackage{graphicx}
\usepackage{bm}
\usepackage{pdfpages}
\input{epsf}
\begin{document}
\epsfverbosetrue
\def\la{{\langle}}
\def\ra{{\rangle}}
\def\vep{{\varepsilon}}
\newcommand{\beq}{\begin{equation}}
\newcommand{\eeq}{\end{equation}}
\newcommand{\beqa}{\begin{eqnarray}}
\newcommand{\eeqa}{\end{eqnarray}}
\newcommand{\q}{\quad}
\newcommand{\A}{|\Omega'|}
\newcommand{\AC}{{\it AC}}
\newcommand{\n}{\\ \nonumber}
\newcommand{\om}{\omega}
\newcommand{\Om}{\Omega}
\newcommand{\tild}{\overline}
\newcommand{\E}{\mathcal{E}}
\newcommand{\os}[1]{#1_{\hbox{\scriptsize {osc}}}}
\newcommand{\cn}[1]{#1_{\hbox{\scriptsize{con}}}}
\newcommand{\sy}[1]{#1_{\hbox{\scriptsize{sys}}}}
%\draft

%\title{Loss and recapture of particles by a time dependent trap  near a continuum threshold }
\title{Adiabaticity in a time dependent trap: a passage  near continuum threshold }
\author {D. Sokolovski$^{a,b}$}
\author {M. Pons$^c$}
%\author {J. G. Muga$^{a,d}$}
\affiliation{$^a$ Departamento de Qu\'imica-F\'isica, Universidad del Pa\' is Vasco, UPV/EHU, Leioa, Spain}
\affiliation{$^b$ IKERBASQUE, Basque Foundation for Science, E-48011 Bilbao, Spain}
\affiliation{$^c$ Departmento de F\' isica Aplicada I, EUITMOP, Universidad del Pa\' is Vasco, UPV-EHU, Bilbao, Spain}
%\affiliation{$^d$ Department of Physics, Shanghai University, 200444 Shanghai, China}
\date{\today}

\begin{abstract}
We consider a time dependent trap externally manipulated in such a way that one of its bound states is brought up towards 
the continuum threshold, and then down again. We evaluate the probability $P^{stay}$ for a particle, initially in a
bound state of the trap, to continue in it at the end of the passage. We use the Sturmian representation, whereby the problem is 
reduced to evaluating the reflecting coefficient of an absorbing potential. In the slow passage limit, $P^{stay}$ goes 
to $1$ for a state turning before reaching the continuum threshold, and vanishes if the bound state crosses into the continuum.
For a slowly moving state just "touching" the threshold $P^{stay}$ tends to a universal value of about $38\%$, for a broad class
of potentials. In the rapid passage limit, $P^{stay}$ depends on the choice of the potential. Various types of trapping potentials are considered, with an analytical solution obtained in the special case of a zero-range well.
\end{abstract}
%\date{\today}
\maketitle
\vskip0.5cm
            % pls. do not remove this line

\section{Introduction}
Recent technological developments have renewed the interest in the dynamics of  a particle, or particles, trapped in bound states of time dependent potentials. External manipulation of Hamiltonians with both discrete and continuum spectra routinely occur in applications  such as metrology and quantum information processing. The presence of a continuum plays an important role in atom lasers \cite{David0,TRAP1}, in the preparation of atomic pulses with a known velocity distribution \cite{velocity}, or in the production of few-body number states \cite{Mark1,Mark2,Fock1,Fock2,Fock3}. Quite often a continuum is responsible for undesirable loss of trapped particles, as it happens in transport of trapped ions, or in trapped ion atomic clocks. An obvious way to avoid such loss is to manipulate the trapping potential sufficiently slowly (adabatically) so that the trapped particle would remain trapped throughout the evolution.
 
The question of adiabaticity in bound-to-continuum transitions, studied by various authors,  \cite{Moy,Kohn,DO,SIEGN}
%\cite{Kohn},\cite{DO},\cite{,
leaves room for further discussion, even in regard to its formulation. 
%Clearly, the standard Landau-Zener  \cite{Land} picture, in which two discrete levels cross, since a continuum cannot be "crossed" in the same sense. 
As a trapping potential becomes shallower, a bound states is brought closer to the continuum, and eventually joins it. With this drastic reorganisation of the adiabatic spectrum, application of methods developed for levels crossing situation, such as the original Landau-Zener model \cite{Land} and its numerous generalisations, is at best problematic. Moreover, in an experimental situation one is likely to control the shape of the trap, 
%rather than directly the energy of the bound state, whose 
so that the evolution of the energy of the bound state near the continuum threshold, must be deduced from that of the potential. With this in mind, one may be interested in asking two distinct questions. Firstly, let the depth of the trap decrease linearly in time. When the evolution stops, what is the probability to remain in the modified bound state? Secondly, let the depth of the potential first decrease, and then increase again, e.g.,  being a quadratic function of time.
%first bringing a bound state closer to the continuum, and then taking it away again. 
What is the probability to remain in a bound state at the end of the passage? The first case was studied in  \cite{DMG}.
In this paper, we consider the second generic case, where a time dependent trap is manipulated in such manner, that a bound state completes a passage near the continuum threshold, first rising towards it, and then moving away again. 
There are three possibilities: the state may "turn" and begin the downward part of its journey before reaching the threshold. Alternatively, it can just "touch" the threshold once, 
or cross into the continuum temporarily, to reappear at a later time. In all cases we will want to know the probability for remaining in the initial state, or, more generlally,  inside the well, once the passage is completed. 

As in  \cite{DMG} we will employ the Sturmian technique, developed in Refs. \cite{ST1,ST2,ST3,ST4} for applications in the theory of atomic collisions. In this way, we reduce the problem of solving a time-dependent Schroedinger equation, to a simpler problem of determining the reflection coefficient of a complex valued "potential". This, in turn, will allow us gain further insight into what happens near  a continuum threshold, and occasionally obtain an exact analytical solution to the problem.

The rest of the paper is organised as follows: is Sect. II we will formulate the problem of a time dependent trap, which can lose a previously bound particle to the continuum. In Sect. III we introduce the Sturmian basis, and use it to expand the particle's state.
In Sect. VI we consider a zero-range  well, and formulate the adiabatic condition for the passage. In Sect. V we solve the zero-range problem exactly for the case where the bound state just touches the continuum threshold. We will show that the probability to remain in the well is independent of the rate of change of the potential, and always equals approximately $38\%$. In Sect. VI the general case for a zero-range potential is analysed. In Sect. VII we consider the Sturmian representation for a rectangular potential, and the corresponding adiabatic limit. In Sect. VIII we employ the single-Sturmian approximation in order to describe the particle's evolution in a rectangular well. In Sect. IX we show that the $38\%$ rule of Sect. V applies universally in the slow passage limit to a wide class of potentials whose evolution is quadratic in time.
Sect. X contains our conclusions.

\section {Loss and recapture of particles by a time dependent potential well}
We start by considering  a particle of a mass $\mu$ in a one-dimensional potential,
\begin{eqnarray}\label{1}
V(x,t)=
%-\Om(t)W(x),\q \Om (t)=
-W(x) \sum_{k=0}^K V^{(k)}t^k, 
\end{eqnarray}
where $W(x)$ is normalised by the condition $\int_{-\infty}^\infty W(x)dx=1$.
The potential is obtained by varying the magnitude of  a finite-range potential well $-W(x)<0$ by means of a time dependent factor,  so that whenever $\sum_{k=0}^K V^{(k)}t^k$ turns negative, $V(x,t)$ becomes a barrier, which doesn't support bound states.
The question we will ask is the following one: if  a particle is put into one of the bound states of the well, $\phi_n$, what is 
the probability to still find it there at some time in the future? The Schroedinger equation (SE) to be solved has the form (we will use $\hbar=1$)
\begin{eqnarray}\label{2}
i\partial_t\Psi(x,t)=-\partial^2_x\Psi/2\mu+V(x,t)\Psi,
\end{eqnarray}
%and needs to be solved with a suitable initial condition, which we will specify shortly. 
%Depending on the sign of $V^{(0)}$, the potential shape at 
%t=0$ may be a well, if $V^{(0)<0$, a barrier, if $\Om_0>0$, or correspond to free motion, if $\Om_0=0$. The particle, prepared initially in $\phi_n$, may remain in it, be transferred to other bound states of the deepening well, or end up ejected into the continuum.
%We are interested in the probabilities of these three outcomes.
and we will assume that  the potential is a deep well in the distant past and future, $V(x,t) <0$ for $t\to\pm \infty$.
%We can specify the initial condition for Eq.(\ref{2})
%still need to specify the initial condition 
%for the SE (\ref{2})
%, and will do it 
%by noting that a 
 A particle in a bound state $\phi_m(x,t)$, 
$\la\phi_m|\phi_m\ra=1$
%, of a deep well 
with a large negative energy $E_m(t) <0$ should continue in it for some time, before approaching the continuum threshold \cite{DMG}. For $\Psi(x,t)$ in Eq.(\ref{a1}) we, therefore, write
\begin{eqnarray}\label{a1}
\lim_{t\to-\infty}\Psi(x,t) = \exp[-i\int^tE_m(t')dt']\phi_n(x,t), 
\end{eqnarray}
%wheis an instantaneous (adiabatic) eigenstate of $V(x,t)$ at a time $t$. 
Similarly, for $t\to \infty$, we should have 
\begin{eqnarray}\label{a2}
\lim_{t\to\infty}\Psi(x,t) =\q \q\q\q\q\q\q\q \q\q\q\q\q\q\q\n
 \sum_n A_{mn}\exp[-i\int^tE_n(t')dt']\phi_n(x,t) +\delta \Psi(x,t), 
\end{eqnarray}
where the first term corresponds to the particles which remained in the well, although possibly not in the same state, 
and $\delta \Psi$ describes the particles lost to the continuum during the passage. 
%%%%%%%%%%%%%%%%%%%%%%%%%%%%%%%%%%%%%%%%%%
\begin{figure}
	\centering
		\includegraphics[width=6cm,height=4cm]{{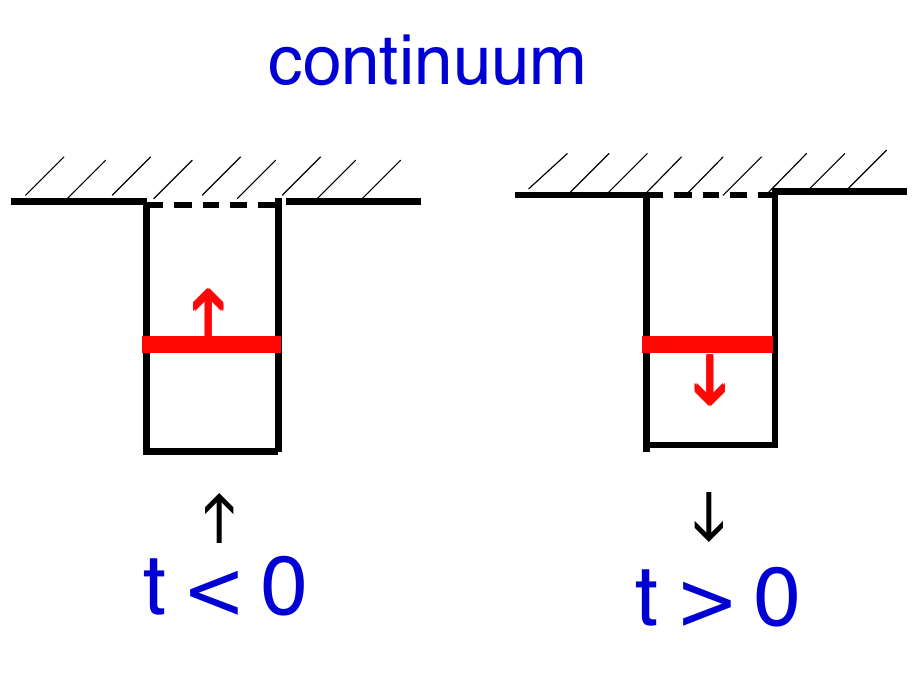}}
\caption{(Color online) Schematic diagram showing the evolution of the potential well described by Eq.(\ref{3a}).
At first the well is becoming shallower, thus bringing its bound state (thick solid) closer to the continuum. Later the well 
deepens, bringing the state down, and possibly bringing more bound states into the well. At $t=0$, the bound state may still exist, 
if $\E<0$, "touch" the continuum threshold, if $\E=0$, a disappear for a while, if $E>0$.
}
\label{fig:3}
\end{figure}Thus, the total probability 
for the particle to stay in the well is given by 
\begin{eqnarray}\label{a3}
P_m^{stay}\equiv\sum_n P^{stay}_{mn}=\sum_n|A_{mn}|^2.
\end{eqnarray}
In the following will will consider the simplest case of  a passage, which is quadratic in time, 
\begin{eqnarray}\label{3a}
 V(x,t)= (\E-v^2 t^2)W(x),
\end{eqnarray}
and of a particle trapped in an ascending  bound state in the distant past, which may remain trapped in one of the descending states, 
or be ejected into the continuum as $t\to \infty$ (see Fig.1). In particular, the ground state of the well (which in one dimension exists as long $V(x,t)<0$ \cite{Land})
will "turn" before reaching the continuum threshold if $\E<0$, "just touch" it if $\E=0$, or, for $\E>0$, disappear at $t=-\sqrt{\E}/v$, before re-appearing again at $t=\sqrt{\E}/v$.  In all three cases, we will be interested in the probabilities $P^{stay}_{mn}$ defined in Eq.(\ref{a3}).
%%%%%%%%%%%%%%%%%%%%%%%%%%%%%%%%%%%%%%%%%%%%%%
\section {Sturmian expansion of the time dependent state}
% $\Psi(x,t=0)=\phi_m(t)$, where $\phi_m(t)$ is the $m$-bound state in the well at
%$t=0$. 
With the help of the Fourier transform, 
\begin{eqnarray}\label{2a}
\Psi(x,t) = \int d\om \exp(-i\om t) \Psi(x,\om), 
\end{eqnarray}
we rewrite Eq.(\ref{2}) as
\begin{equation}\label{3}
\om \Psi(x,\om) = -\partial^2_x\Psi/2\mu-W(x)\sum_{k=0}^K(-i)^kV^{(k)}\partial^k_\om\Psi.
\end{equation}
and look for a suitable basis in which  to expand $\Psi(x,\om)$. Using the set of the positive-energy scattering states describing particles incident on $V(x,t)$ from left and right is one option, yet there is a more convenient one. 
Particles ejected from the well should be described by outgoing waves on both sides of the potential. 
%With no source at infinity, the only particles travelling on both sides of $V(x,t)$ will be those ejected from the well itself. Such particles should be described by outgoing waves on both sides of the potential.
Sturmian basis sets with the desired properties are well known in literature \cite{ST3}. 
They are obtained by imposing outgoing boundary conditions, fixing the value of $\om$ in Eq.(\ref{3}), and searching for particular shapes of $V_n(x,t)=\rho_n W(x)$, $n=1,2,..$ such that the stationary SE
\begin{eqnarray}\label{4}
-\partial^2_xS_n/2\mu+\rho_nW(x)S_n=\om S_n, \q n=0,1,2...
\end{eqnarray}
has  a solution $S_n(x,\om)$ which satisfies the boundary conditions 
\begin{eqnarray}\label{5}
S_n(x,\om)\sim \exp(\pm i\sqrt{2\mu\om}x), \q x\to \pm \infty.
\end{eqnarray}
The Sturmian eigenfunctions $S_n$ (also known as Sturmians) differ for positive and negative $\om$'s. As seen from Eq.(\ref{5}), for $\om < 0$, all $S_n(x)$ exponentially decay on both sides of the well, $S_n(x)\sim \exp(-\sqrt{2\mu|\om|}|x|)$ for $|x|\to \infty$,
%and are just the bound states of $\rho_n W(x)$, $\rho_n<0$.  
%A very deep potential well may support a number of bound states which, as the well gets shallower, move upwards and eventually join the continuum.
%Discrete real Sturmian eigenvalues $\rho_n(\om)$ select, therefore,  potential shapes, such 
so that $\rho_n W(x)$ has a bound state at the chosen energy $\om$ \cite{FOOT}.
%\newline
For $\om>0$, a Sturmian contain outgoing travelling waves, $S_n(x,\om) \sim \exp(\pm i\sqrt{2\mu\om}|x|)$ as $x\to\pm \infty$.
% on both sides of the well. 
 This can only be the case if $\rho_n W(x)$ is a complex valued emitting potential, which, in turn,  requires $Im \rho_n>0$ for $\om>0$. In general, as $\om$ changes from $-\infty$ to $+\infty$, a chosen $\rho_n(\om)$ traces a continuous trajectory in the complex $\rho$-plane. 
 \newline
From 
Eq.(\ref{4}) follows an orthogonality relation
\begin{eqnarray}\label{5a}
(S_m(\om)|S_n(\om))\equiv\int S_m(x,\om)W(x)S_m(x,\om)dx=\n
\delta_{mn}\times(S_n(\om)|S_n(\om)),
\end{eqnarray}
where $\delta_{mn}$ is the Kroneker delta.
The Sturmians are also known to form complete sets (we refer the reader to Ref.\cite{ST3} for a detailed discussion). Thus, to construct a physical solution $\Psi(x,t)$ describing particles which escape from the trap,
%containing as $|x|\to \infty$  only decaying and outgoing ways, 
we expand $\Psi(x,\om)$ in (\ref{3}) in the basis of $S_n$,
\begin{eqnarray}\label{6}
 \Psi(x,\om)=\sum_{n}B_n(\om)S_n(x,\om),
\end{eqnarray} 
where the coefficients $B_n(\om)$ are to be determined. Inserting (\ref{6}) into Eq.(\ref{3}), after adding and subtracting $\sum_{n}\rho_nW B_nS_n$,  we have
\begin{eqnarray}\label{7}
%\om \Psi(x,\om) = -\partial^2_x\Psi/2+\sum_{k,m}(-i\om)^k\Om_k\partial^k_\om[B_n(\om)S_n(x,\om)]
\sum_n\{\sum_k(-i)^kV^{(k)}W(x)\partial^k_\om[B_n(\om)S_n(x,\om)]+\n
\rho_nW(x)B_n(\om)S_n(x,\om)\}=0.\q\q
\end{eqnarray}
In our quadratic case (\ref{3a}), multiplication of Eq.(\ref{7}) by $S_m(x,\om)$ and integration over $x$, yields the following set of equations for $B_n(\om)$, 
\begin{eqnarray}\label{8}
%\nonumber
M^{(0)}_{mm}[v^2B_m''+(\E-\rho_m)B_m]-
\q\q\q\q\q\q\n
v^2\sum_n (2M^{(1)}_{mn}B_n'+M^{(2)}_{mn}B_n)=0,
%\\
%+\Om_0\sum_n M^{(0)}_{mn}B_n-\rho_mB_m(S_m(\om)|S_m(\om)),\q\q
\end{eqnarray}
where a prime denotes differentiation with respect to $\om$, and 
\begin{eqnarray}\label{9}
M^{(j)}_{mn}\equiv\int S_m(x,\om)W(x)\partial_\om^jS_n(x,\om)dx\equiv\n
(S_m(\om)|S^{(j)}_n(\om)).\q\q
\end{eqnarray}
Equations (\ref{8}) and (\ref{9}) are the main achievement of the Sturmian approach: the problem of solving  a partial differential equation (\ref{2}) is reduced to solving a system of second-order ordinary differential equations.
% thus avoiding the use of continuum scattering states.
With only few terms in Eqs.(\ref{8}) usually needed, the Sturmian approach offers a significant computational advantage in case of many dimensions \cite{ST1,ST2,ST3,ST4}.  In the one-dimensional case considered here, it can offer a further insight into the physics of scattering by time-dependent potentials and simplify calculations in certain limiting cases, as we will demonstrate next.
%%%%%%%%%%%%%%%%%%%%%%%%%%%%%%%%%%%%%
%\section{A zero-range well}
\section{quadratic zero-range model. The adiabatic limit}
%The initial condition (\ref{a1}) needs to be translated into an initial condition for the system of equations (\ref{7}), and we will first do it 
%for
We start with the simplest case of 
 a zero-range potential,
\begin{eqnarray}\label{a1a} 
W(x)=\delta(x),
\end{eqnarray} 
which for $\E-v^2 t^2 < 0$ support a single adiabatic bound state ($\theta(x)=1$ for $x\ge 0$ and $0$ otherwise)
\begin{eqnarray}\label{a1b} 
\phi_0(x,t)=[-2\mu E_0(t)]^{1/4}[\theta(x) \exp(i\sqrt{2\mu E_0(t)}x)+
%\q\q\q\q 
%\q\q\q\q\q
\q\q\n
\theta(-x)\exp(-i\sqrt{2\mu E_0(t)}x)],
%/\sqrt{-2\mu E(t)}
\end{eqnarray}
with an energy 
\begin{eqnarray}\label{a1c} 
E_0(t)=-\mu(\E-v^2 t^2)^2/2.
\end{eqnarray}
%\section{the quadratic zero-range model}
There is a only one Sturmian \cite{DMG}, \cite{ST4} 
%(we omit the subscript $n$)
\begin{equation}\label{a4}
S_0(x,\om)=[\theta(x) \exp(i\sqrt{2\mu\om}x)+\theta(-x)\exp(-i\sqrt{2\mu\om}x)],
\end{equation}
and the corresponding Sturmian eigenvalue, given by
\begin{equation}\label{a5}
\rho_0(\om)=i\sqrt{2\om/\mu},
\end{equation}
is single-valued on a two-sheet Riemann surface $\mathcal{R}$ of  $\sqrt{\om}$, cut along the positive semi-axis. With no other Sturmians present, 
and $M_{00}^{(2)}=0$ since
%, according to Eq.(\ref{a4}),
 $S(0,\om)=1$, 
%Eqs.(\ref{8}) reduce to 
taking complex conjugate of Eqs.(\ref{8}) \cite{CONJ} yields
\begin{eqnarray}\label{a6}
%\nonumber
{B_0^*}''+v^{-2}q^2B_0^*=0, \q q(\om)\equiv \sqrt{\E-\rho_0^*(\om)}
%\frac{1}{v}\left[ \frac{-\Om_0}{2}+\frac{\rho}{2}\right ]^{1/2}.
\q\q\q\q\q\q\n
%\\
%+\Om_0\sum_n M^{(0)}_{mn}B_n-\rho_mB_m(S_m(\om)|S_m(\om)),\q\q
\end{eqnarray}
must be integrated along the contour running along the real $\om$-axis above the cut the first sheet of $\mathcal{R}$, where $S_0(x,\om)$ satisfies the required outgoing/decaying waves boundary conditions (\ref{5}).
\newline
Equation (\ref{a6}), which is exact, can now be read in a completely different manner. It has the form of a stationary SE describing a "particle" of a "mass" $1/2$ with a 
"coordinate" $\om$, 
of an "energy" $\mathcal{E}$,  scattered by a "potential" $\mathcal{W}(\om)=\rho_0^*(\om)$, with $v$ playing the of the "Planck constant" $\hbar$ \cite{FOOT1}. [We will always use the quotes when we refer to the fictitious "particle" in Eq.(\ref{a6}), in order to distinguish it from the real particle described by the SE (\ref{2}).]
%\newline
% , but first we need to specify a contour on $\mathcal{R}$ along which to integrate. 
%A quick inspection reveals that Eq.(\ref{a6}) 
We note that  the "potential" $\mathcal{W}$, shown in Fig. 2,  has a valley ($Re\mathcal{W}<0$, $Im\mathcal{W}=0$) for $\om<0$, and becomes purely absorbing, 
($Re\mathcal{W}=0$, $Im\mathcal{W}<0$) 
 for $\om>0$.
\newline
Properties of equations of the type (\ref{a6}) are well known (see, e.g., \cite{Land}). As $\om \to -\infty$, $q(\om)\to \infty$, while $\mathcal{W}(\om)$ becomes flatter,
$\mathcal{W}'(\om)\sim 1/ \sqrt{|\om|}\to 0$, so that $B(\om)$ can be expressed in semiclassical form \cite{Land}, in terms of "incoming" (+) and "outgoing" (-) "waves",   
\begin{eqnarray}\label{a7}
%\nonumber
B_0^*(\om) \approx \frac{A^+}{\sqrt{q(\om)}}\exp[\frac{i}{v}\int^\om q(\om')d\om']+\q\q\q\q\q\q\n
\frac{A^-}{\sqrt{q(\om)}}\exp[-\frac{i}{v}\int^\om q(\om')d\om'],\q \om \to -\infty\q
\end{eqnarray}
where $A_{\pm}$ are unknown constants to be determined. We do not expect the particle to acquire very high energy, and must, therefore, require that $B(\om)\to 0$ as $\om\to \infty$. Thus, taking the principal branch of the square root in Eq.(\ref{a6}), we have
\begin{eqnarray}\label{a8}
%\nonumber
B^*(\om) \sim\frac{1}{\sqrt{q(\om)}}\exp[-\frac{1}{v}\int^\om |q(\om')|d\om']\to 0, \q \om \to \infty.\q\q
%\frac{A-}{\sqrt{q(\om)}}\exp[-i\int^\om q(\om')d\om']\q
\end{eqnarray}

Finally, inserting (\ref{6}) and (\ref{a7}) into Eq.(\ref{2a}) we note that as $t\to \pm\infty$ the integral over $\om$ may be evaluated by the stationary phase method \cite{BRINK}, 
%The phase of the integrand is stationary at $\om=\om_s$,  such that
%\begin{eqnarray}\label{a9}
%t=\pm q(\om_s),
%\end{eqnarray}
and as $t\to -\infty$ we have (details are given in the Appendix)
\begin{eqnarray}\label{a9}
\Psi(x,t) \approx 2v\sqrt{\pi i}A^+\phi(x,t) \exp[-i\int^t E(t')dt'].
\end{eqnarray}
This describes a particle trapped in the ascending bound state, which approaches the continuum 
threshold from below. Similarly, for $t\to \infty$ we have 
\begin{eqnarray}\label{a10}
\Psi(x,t) \approx 2v\sqrt{-\pi i}A^-\phi(x,t) \exp[-i\int^t E(t')dt']\n
 + \delta \Psi(x,t),\q\q
\end{eqnarray}
where 
%the first term is the contribution from the stationary region around $\om_s$, and $ \delta \Psi(x,t)$ contains the remaining part of the integral. The 
first term describes a particle trapped in the descending bound state, moving away from the continuum threshold.
For the probability to complete the passage, and remain in the bound state we, therefore, have 
\begin{eqnarray}\label{a11}
P_{00}^{stay}=\frac{|A_-|^2}{|A_+|^2},
\end{eqnarray}
which is less or equal to one, as guaranteed by the absorbing nature of the "potential" for $x>0$.
%since there are always more "particles" moving towards the absorbing "potential". 
\newline
Reduction of the original time dependent problem to the one of determining the reflection coefficient of a 
complex valued barrier allows us to prove the existence of the adiabatic limit in case the bound state "turns" without touching the continuum $\E<0$.  Now "absorption" represents the loss of the particle to the continuum, and to be absorbed, 
the "particle" must first cross the "classically forbidden region" (see Fig. 2), impenetrable in the "classical limit" $v\to 0$.
This is the {\it adiabatic theorem}. The behaviour for $\E\ge 0$ requires somewhat more attention, and we will consider it next.

%%%%%%%%%%%%%%%%%%%%%%%%%%%%%%%%%%%%%%%%%%%%%%
\section{A zero-range well: just touching the continuum}
 %%%%%%%%%%%%%%%%%%%%%%%%%%%%%%%%%
\begin{figure}
	\centering
		\includegraphics[width=8cm,height=8cm]{{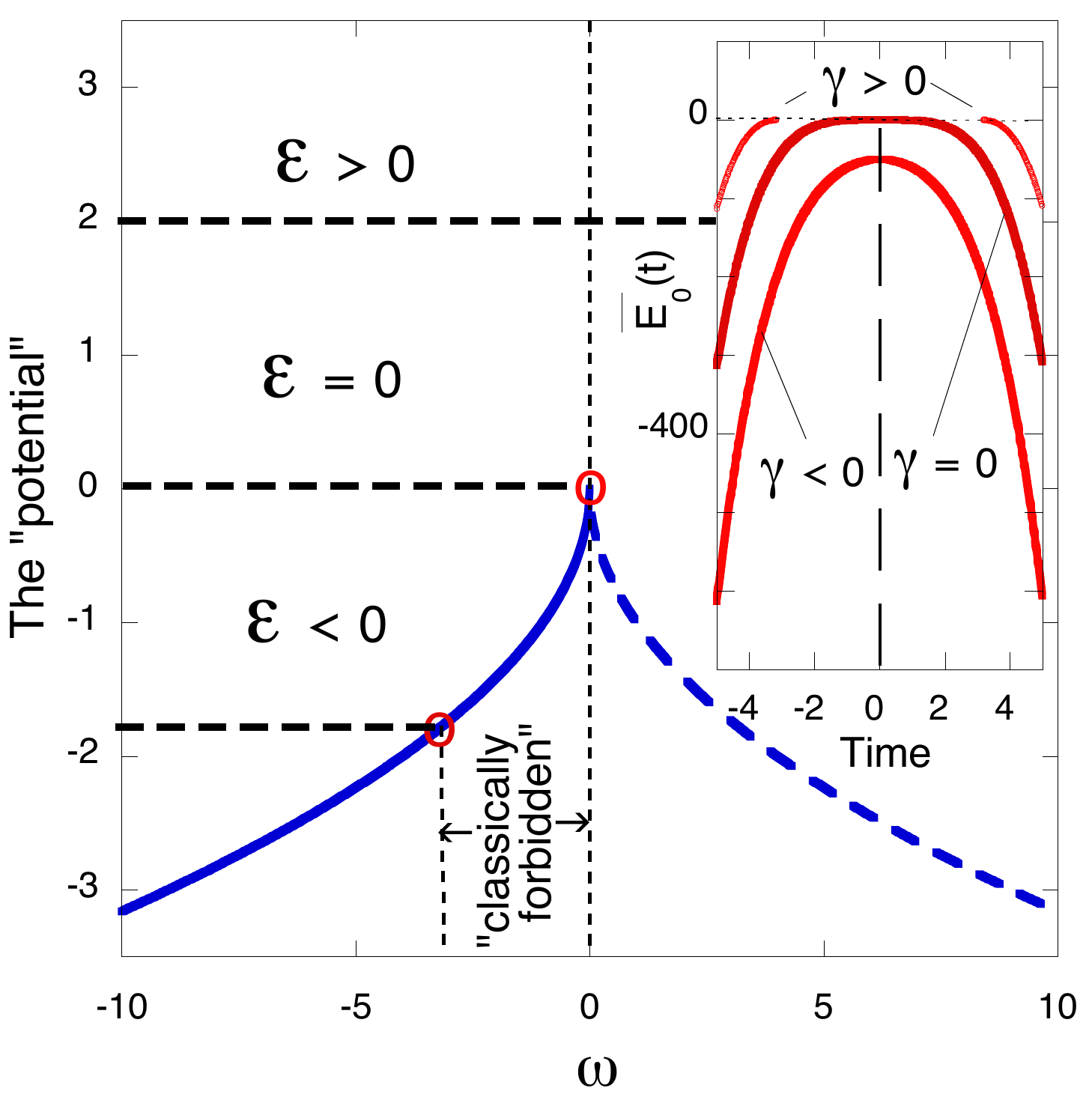}}
\caption{(Color online). Schematic diagram for the zero-range model.  A "particle" with an "energy" $\E$ is scattered by a complex-valued "potential", 
$\mathcal{W}$, whose real and imaginary parts are shown by the solid and dashed lines, respectively. Also indicated is the "classically forbidden region", separating a "particle" with $\E <0$ from the "absorption" region $\om>0$. The inset shows the energy of the bound state given by (\ref{a1c}), $\overline{E}_0(t)=\mu^{-1/5}v^{-4/5}E_0(t)$, as a function of. $\overline{t}=\mu^{1/5}v^{4/5}t$.}
\label{fig:3}
\end{figure}
With $\E=0$, we have $E_0(t=0)=0$, so the bound state of a zero-range well approaches the continuum threshold, and "touches" at the moment it "turns" to begin the downward leg of its journey. In this special case the equation for $B_0$, 
 \begin{eqnarray}\label{b1}
%\nonumber
B_0''+b\sqrt{\om}B_0=0,\q b=\sqrt{2/\mu}\exp(3\pi i/2)/v^2,
\end{eqnarray}
can be solved analytically in terms of the Bessel functions \cite{GRAD}. The solution which vanishes as $\om\to \infty$,  
%on the contour $\Gamma$ shown in Fig. , For $\om > 0$ it is given by 
 is given by
 \begin{eqnarray}\label{b2}
%\nonumber
B(\om)=\sqrt{\om}H^{(1)}_{2/5}(z), \q z\equiv \frac{2^{9/4}}{5v}e^{3\pi i/4}\om^{5/4},
\end{eqnarray}
where $H^{(j)}_{\nu}(z)$, $j=1,2$ is the Hankel function of the $j$-th kind \cite{WATS}. As $\om \to \infty$, we have (omitting inessential phase factors)
 \begin{eqnarray}\label{b3}
%\nonumber
%B(\om)\sim \om^{-1/8}\exp[iA\om^{5/4}-A\om^{5/4}]\to_{\om\to \infty}0,\q
B(\om)\sim \om^{-1/8}\exp[-(1+i)K\om^{5/4}]_{\om\to \infty}\to0,
\end{eqnarray}
where $K\equiv 2^{7/4}/5\mu^{1/4}>0$. Equation (\ref{b3}) is readily recognised as a special case of Eq.(\ref{a8}), with $q(\om)=(2\om/\mu)^{1/4}\exp(-i\pi/4)$. To find the asymptotic form of $B(\om)$ for $\om \to -\infty$, we use the formula connecting the values of 
$H^{(1)}_{2/5}(z)$ on the ray $z'=\rho\exp(3\pi i/4)$ with those along $z=z'\exp(i\pi)=\rho\exp(7\pi i/4)$ (see \cite{WATS}, Sect. 3.62).
 \begin{eqnarray}\label{b4}
%\nonumber
H^{(1)}_{2/5}(z')=2\cos(2\pi/5)H^{(1)}_{2/5}(z)
%\n
+\exp(-2\pi i/5)H^{(2)}_{2/5}(z).\q\q
\end{eqnarray}
Recalling that  $H_\nu^{(1,2)}(z) \sim (2/\pi z)^{1/2}\exp[\pm i(z-\nu\pi/2-\pi/2)]$ \cite{WATS}, 
and taking complex conjugate of Eq.(\ref{b4}),
 we identify $H^{(2)}_{2/5}(z)$ with the "incoming wave" in Eq.(\ref{a7}), which gives 
 \begin{eqnarray}\label{b5}
%\nonumber
P_{00}^{stay}(v)=4\cos^2(2\pi/5) \approx 0.38197.
\end{eqnarray}
Thus, for a narrow well such that its bound state just "touches" the continuum at $t=0$, the probability to remain in the well 
is independent of the rate of change of the potential. There is a perfect balance:  a rapidly changing well is more likely to eject the particle  into the continuum, yet the time the bound state spend near the threshold is short. If the well changes slowly, this time is longer, yet the particle is ejected less efficiently. As a result, there is no adiabatic limit as $v\to 0$, and the value of $P_{00}^{stay}$ is always given by Eq.(\ref{b5}). Below we will show that Eq.(\ref{b5}) has a more general meaning, also beyond the zero-range model considered in this Section.
%%%%%%%%%%%%%%%%%%%%%%%%%%%%%%%%%%%%%%%%%%%%%%%%%%%%%%%%%%
\section{A zero-range well: the general case}
No analytic solution of  (\ref{b1}) is known (at least to us) for $\E \ne 0$, so the equation must be solved numerically. We note first that, for a narrow well (\ref{a1a}), $P^{stay}_{00}$ is determined by a single dimensionless parameter 
 \begin{eqnarray}\label{c1}
%\nonumber
\gamma\equiv \frac{\E\mu^{2/5}}{v^{2/5}}.
\end{eqnarray}
Indeed, in the scaled variables $\tau=\mu^{1/5}v^{4/5}t$ and $y=\mu^{3/5}v^{2/5}x$, the SE (\ref{2}) reads 
$i\partial_{\tau}\Psi(y,\tau)=-\partial^2_y\Psi/2+(\gamma-\tau^2)\delta(y)\Psi$, and Eq.(\ref{b1}) only needs to be 
solved for $v=1$, and various values of $\E=\gamma$. 
The dependence of $P_{00}^{stay}(\gamma)$ on $\gamma$ is shown in Fig. 3. 
The probability tends to $1$ for $\gamma \to -\infty$, where the "absorbing potential" in Fig. 2 is separated by a broad "classically forbidden" region.
At $\gamma=0$ the curve passes through the value given by Eq.(\ref{b5}), $P^{stay}(0)=4\cos^2(2\pi/5)$, and tends to zero as $\gamma \to \infty$, i.e., when the "particle" can penetrate deep into the "absorbing region", and nothing is "reflected". 
\newline
Using Figure 3, it is easy to predict the behaviour of the retention probability $P^{stay}$ as a function of $v$, for a given $\E$.
For $\E < 0$, and $v<<\mu |\E|^{5/2}$ the passage will be adiabatic, with almost none of the particles lost.
For $\E > 0$ and $v<<\mu |\E|_0^{5/2}$, the bound state will disappear for a long time  (see inset in Fig. 2), and none of the particles will be recovered when the it finally reappears. With $v\to \infty$, $\gamma$ will vanish for any choice of $\E$, and we have
 \begin{eqnarray}\label{c1a}
lim_{v\to \infty} P_{00}^{stay}(\E,v)=4\cos^2(2\pi/5)
\end{eqnarray}
so that a rapidly changing zero-range well will retain the particle in about $38\%$ of all cases, regardless of the value of $\E$. 
The dependence of $P_{00}^{stay}(\E,v)$ on $v$ for different values of $\E$ is shown in Fig. 4.
\newline
Finally, in order to study the evolution of the population $P_0(t)$ of the moving bound state, we solved numerically the original SE (\ref{2}). The results shown in Fig. 5 demonstrate that $P_0(t)\equiv |\la\phi_0(t)|\Psi(t)\ra|^2$ undergoes oscillations before reaching the asymptotic value  
$P_0(t)=P_{00}^{stay}$, when the bound state is well removed from the continuum.
\begin{figure}
	\centering
		\includegraphics[width=9cm,height=6cm]{{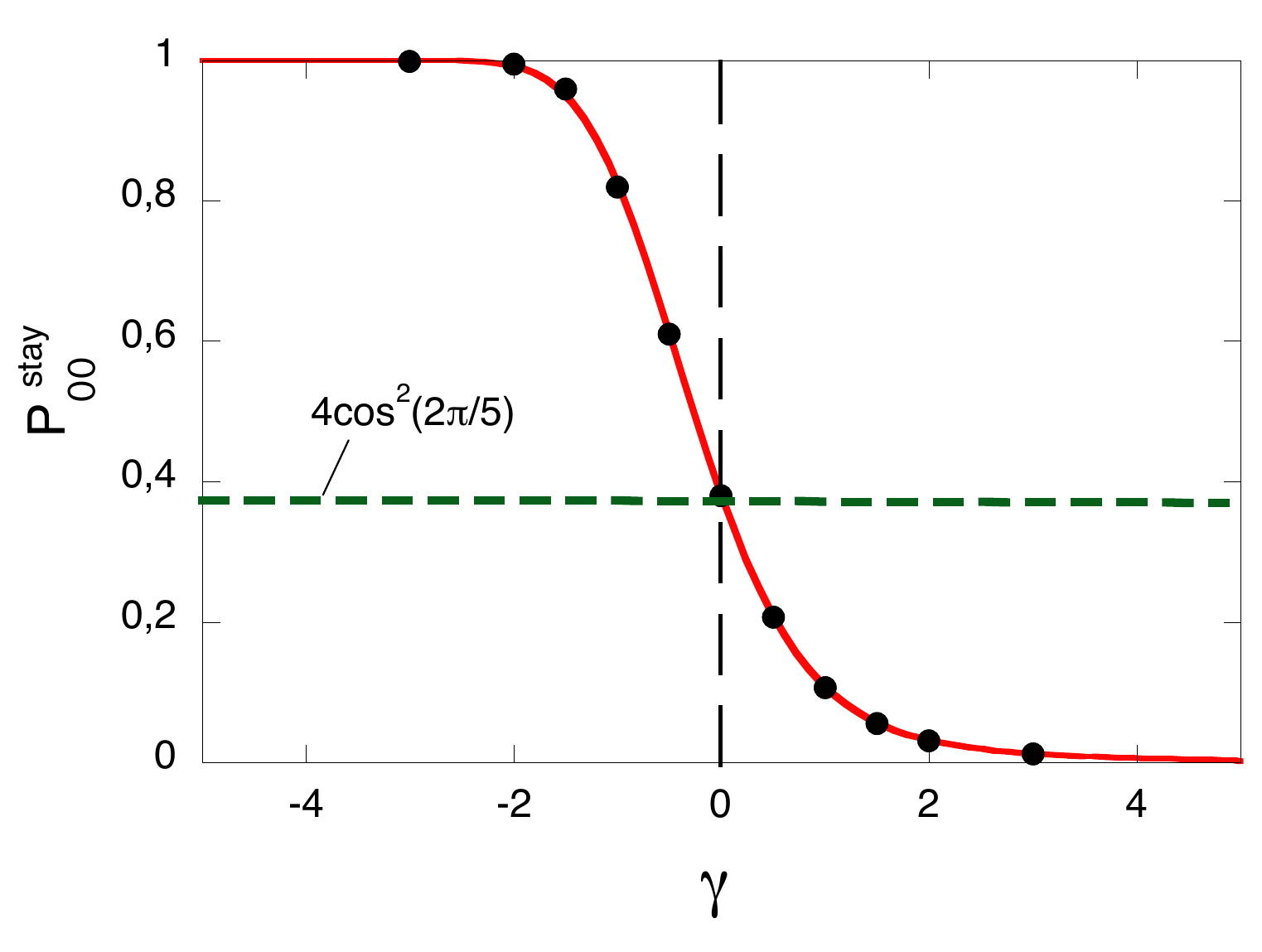}}
\caption{(Color online) The probability $P^{stay}_{00}$ vs. $\gamma$  for the quadratic zero range model (\ref{a1a}),
obtained by integration of Eq.(\ref{a6}) (solid), and by solving numerically the original SE (\ref{2}) (dots).
}
\label{fig:3}
\end{figure}
\begin{figure}
	\centering
		\includegraphics[width=9cm,height=6cm]{{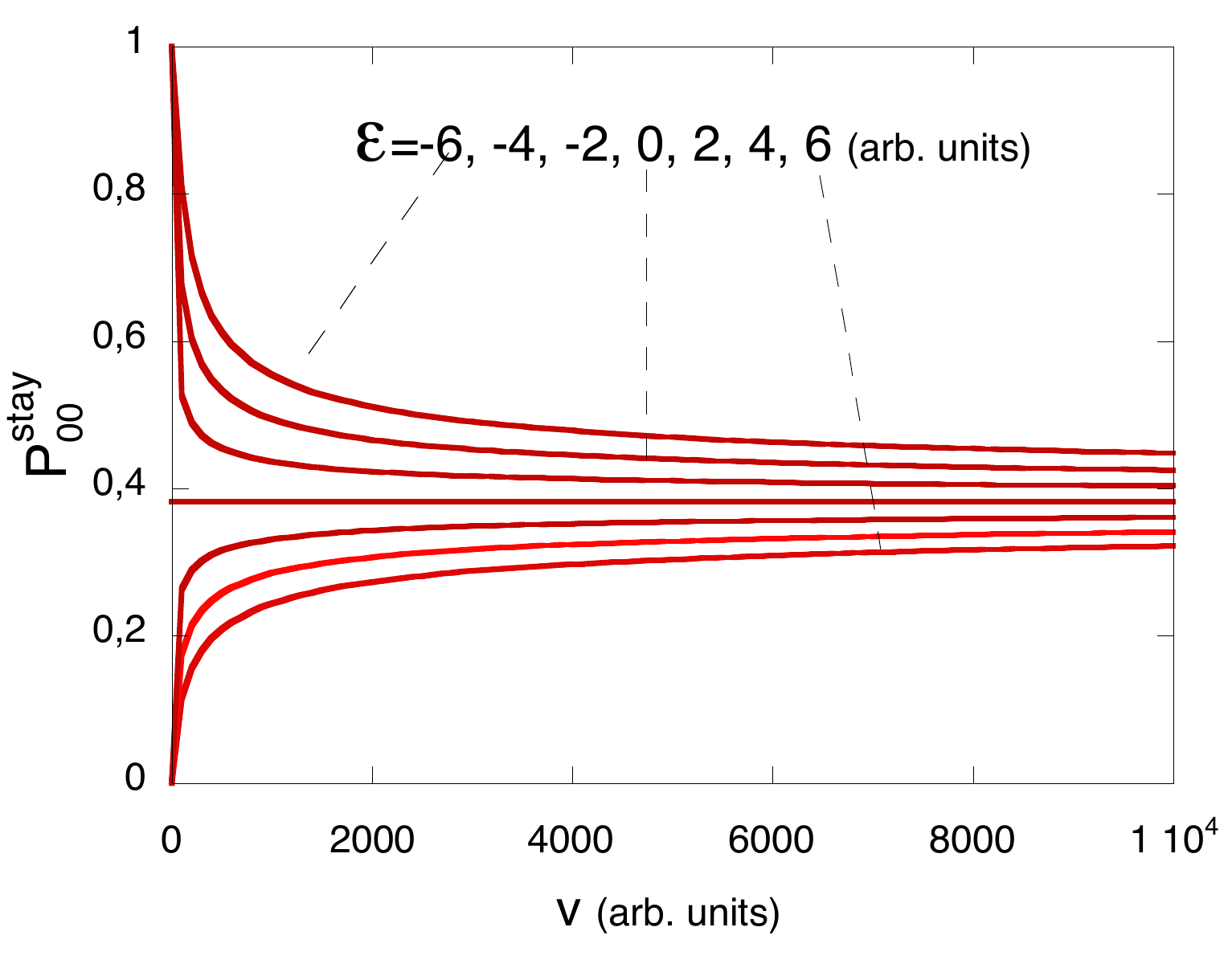}}
\caption{(Color online)The probability $P^{stay}_{00}$ vs. $v$  for the quadratic zero range model (\ref{a1a}), 
for different values of $\E$.
}
\label{fig:3}
\end{figure}
\begin{figure}
	\centering
		\includegraphics[width=9cm,height=6cm]{{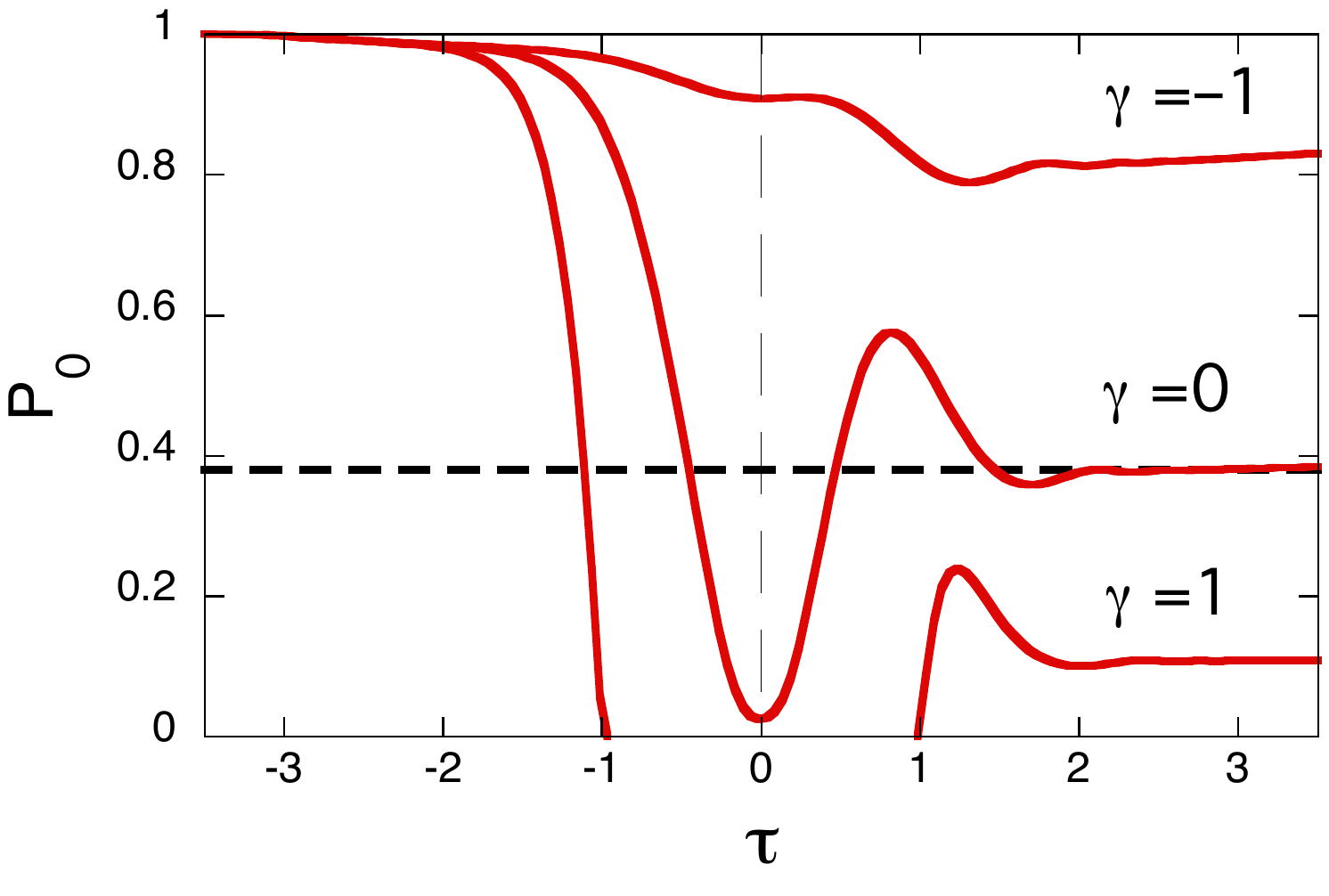}}
\caption{(Color online) The population of the moving bound state, $P_0(t)=|\la\phi_0(t)|\Psi(t)\ra|^2$, vs. $\tau=\mu^{1/5}v^{4/5}t$ for three values
of $\gamma=\E\mu^{2/5}v^{-2/5}$. For $\gamma=1$ the bound state disappears at $\tau=-1$, and reappears again at $\tau=1$.
}
\label{fig:3}
\end{figure}
%%%%%%%%%%%%%%%%%%%%%%%%%%%%%%%%%%%%%%%%%%%%%%%
\section{A rectangular well: the adiabatic limit}
A more realistic case of a rectangular well of a width $2a$,
 \begin{eqnarray}\label{r1}
%\nonumber
W(x)=[\theta(x+a)\theta(a-x)]/2a,
%1\q for \q -a\le x \le a, \q and \q 0 \q otherwise.
\end{eqnarray}
is somewhat more involved. There are two types of Sturmians, symmetric and antisymmetric about the origin, $S_n(x,\om)=S_n(-x,\om)$, and $T_n(x,\om)=-T_n(-x,\om)$. For $-a\le x \le a$, these are given by 
 \begin{eqnarray}\label{r2}
 S_n(x,\om) = \cos(p_nx)/[1+\sin(2p_na)/2p_n]^{1/2},\n 
 \q n=0,2,4...,
\end{eqnarray}
and 
 \begin{eqnarray}\label{r3}
 T_n(x,\om) = \sin(p_nx)/[1-\sin(2p_na)/2p_n]^{1/2},\n
  \q n=1,3,5...
\end{eqnarray}
so that $(S_n|S_n)=(T_n|T_n)=1$.
Since the matrix elements in Eqs.(\ref{8}) couple only Sturmians of the same parity, $(S_m(\om)|T^{(j)}_n(\om))=(T_m(\om)|S^{(j)}_n(\om))=0$, we may limit our analysis to the case where a particle is prepared initially in a bound state symmetric about the origin. The corresponding Sturmian eigenvalues $\rho_n$, $n=0,2,4...$, are then found by solving a transcendental equation, 
 \begin{eqnarray}\label{r4}
\sin(p_na)/\cos(p_na)+ik/p_n=0, 
\end{eqnarray}
where
 \begin{eqnarray}\label{r5}
p_n(\om)= \{2\mu[\om-\rho_n(\om)]\}^{1/2}, \q and \q k(\om)=(2\mu \om)^{1/2}.\q\q
\end{eqnarray}
Thus, $\rho_n(\om)$, is the magnitude of the rectangular potential, real o complex, such that at a given energy $\om$, there is a symmetric solution $S_n(x)$  of the SE (\ref{4}), satisfying the boundary conditions (\ref{5}).
%\cite{FOOTrect}. 
\newline
The interpretation of equations for $B^*_n(\om)$ 
%defined in Eq.(\ref{}) 
is similar to that given in Sect. IV.
One may think of a fictitious "particle" with an "energy" $\E$ which can move on several complex valued "potential surfaces" $\mathcal{W}_n(\om)=\rho_n^*(\om)$.  On each "surface", "absorption", possible for $\om>0$,  accounts for the loss of the real particle to the continuum. There is also a possibility for hopping between the "surfaces", facilitated by matrix elements $M^{(1)}_{mn}(\om)$ and $M^{(2)}_{mn}(\om)$. If a particle is prepared in the $m$-th state of the deep well, we must look for a solution of this "coupled channels problem" 
containing, as $\om \to -\infty$,  an incoming wave on the $m$-th "potential surface" and, possibly, "outgoing waves" in all other "channels",  
%b 
\begin{eqnarray}\label{r6}
B_n^*(\om) \sim \delta_{mn}\frac{A_m^+}{\sqrt{q_m}}\exp[\frac{i}{v}\int^\om q_m(\om')d\om']+\q\q\q\q\q\q\n
\frac{A_n^-}{\sqrt{q_n}}\exp[-\frac{i}{v}\int^\om q_n(\om')d\om'],\q \om \to -\infty,\q \q\q 
\end{eqnarray}
where $m,n=0,2,4...$. The probabilities for a particle to start in the $m$-th, and end up in the $n$-th adiabatic bound states, are given by 
\begin{eqnarray}\label{r7}
P_{mn}^{stay}= \frac{|A^-_n|^2}{|A^+_m|^2},\q m,n =0,2,4...
\end{eqnarray}
\begin{figure}
	\centering
		\includegraphics[width=9cm,height=7cm]{{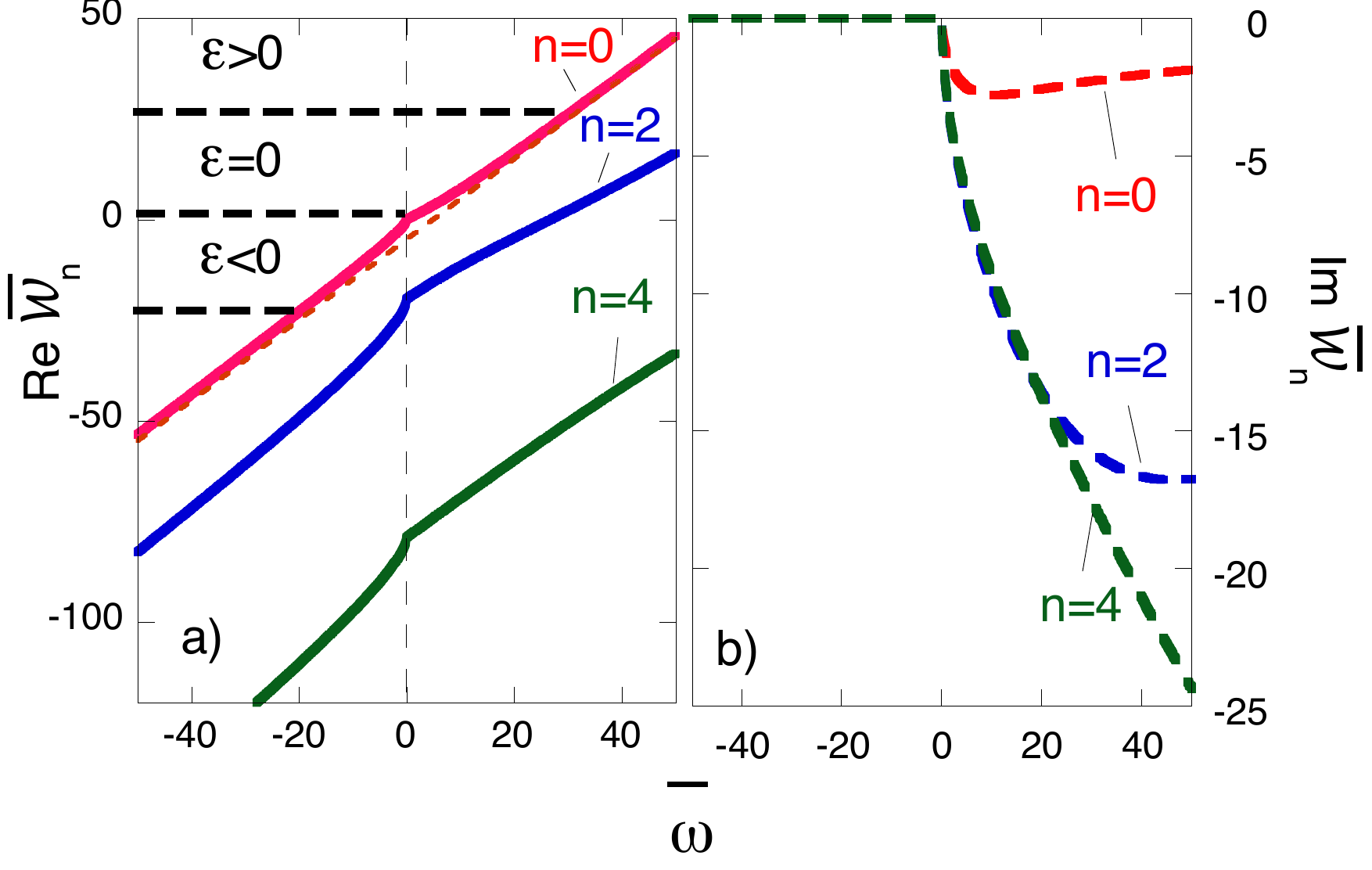}}
\caption{(Color online) Real (a) and imaginary (b) parts of the "channel potentials" $\overline{\mathcal{W}}_n=\mu a \mathcal{W}_n$ vs. $\overline{\om}=\mu a^2 \om$ for $n=0,1,2$. In a), also shown by a dashed line is the large-$\om$ asymptote of $ Re \mathcal{W}_0$, $\om a-\pi^2/2\mu a$.
}
\label{fig:3}
\end{figure}
The "potentials" $\mathcal{W}_n(\om)$ are shown in Fig. 6 for $m=0,2,4$. We have $Im \mathcal{W}_n(\pm)\equiv 0$ for
$\om<0$, where Sturmians are just bound states of a real potential well. We also note that
 \begin{eqnarray}\label{r8}
\lim_{\om\to \pm\infty}Re \mathcal{W}_n(\om)=\om a-(2n+1)^2\pi^2/2\mu a, 
\end{eqnarray}
and 
 \begin{eqnarray}\label{r9}
\lim_{\om\to \infty}Im \mathcal{W}_n(\om)=0. 
\end{eqnarray}
This is because for a large negative $\om$, 
%the $n$-th Sturmian is just the $n$-th eigenstate 
the Sturmians tend to the eigenstates
of a potential box with infinite walls at $x=\pm a$. Since the energy of the state is $\om$, 
%the depth of the well, 
$\rho_n$, is found by subtracting from $\om$ the energy of the $n$-th state, as measured from the floor of the well. In the opposite limit, $\om\to \infty$, the particle becomes bound at the top of an infinitely high rectangular barrier. These bound states, quantised between the sharp potential drops at $x=\pm a$, are essentially the same as those quantised between the walls of an infinite potential box \cite{BR}. Since the Sturmians cease to depend on $\om$ as $\om\to \pm\infty$, 
 \begin{eqnarray}\label{r10}
%\lim_{\om\to \pm\infty} 
S_n(x,\om)\sim \theta(x+a)\theta(a-x)\cos [(2n+1)\pi x/2a], 
\end{eqnarray}
 the matrix elements, coupling the "potential surfaces", vanish in the same limit, 
 \begin{eqnarray}\label{r11}
%\lim_{\om\to \pm\infty} 
M_{mn}^{(1,2)}(\om)\to 0, \q \om\to \pm \infty.
\end{eqnarray}
Both $M_{mn}^{(1)}$ and $M_{mn}^{(2)}$ are singular at the threshold $\om=0$, 
as is explained in the Appendix B. In particular, for $M^{(2)}_{00}$, which will be required
in the next Section, we find $M^{(2)}_{00}(\om) \sim \om^{-1.5}$ (see Fig. 7).
\begin{figure}
	\centering
		\includegraphics[width=9cm,height=7cm]{{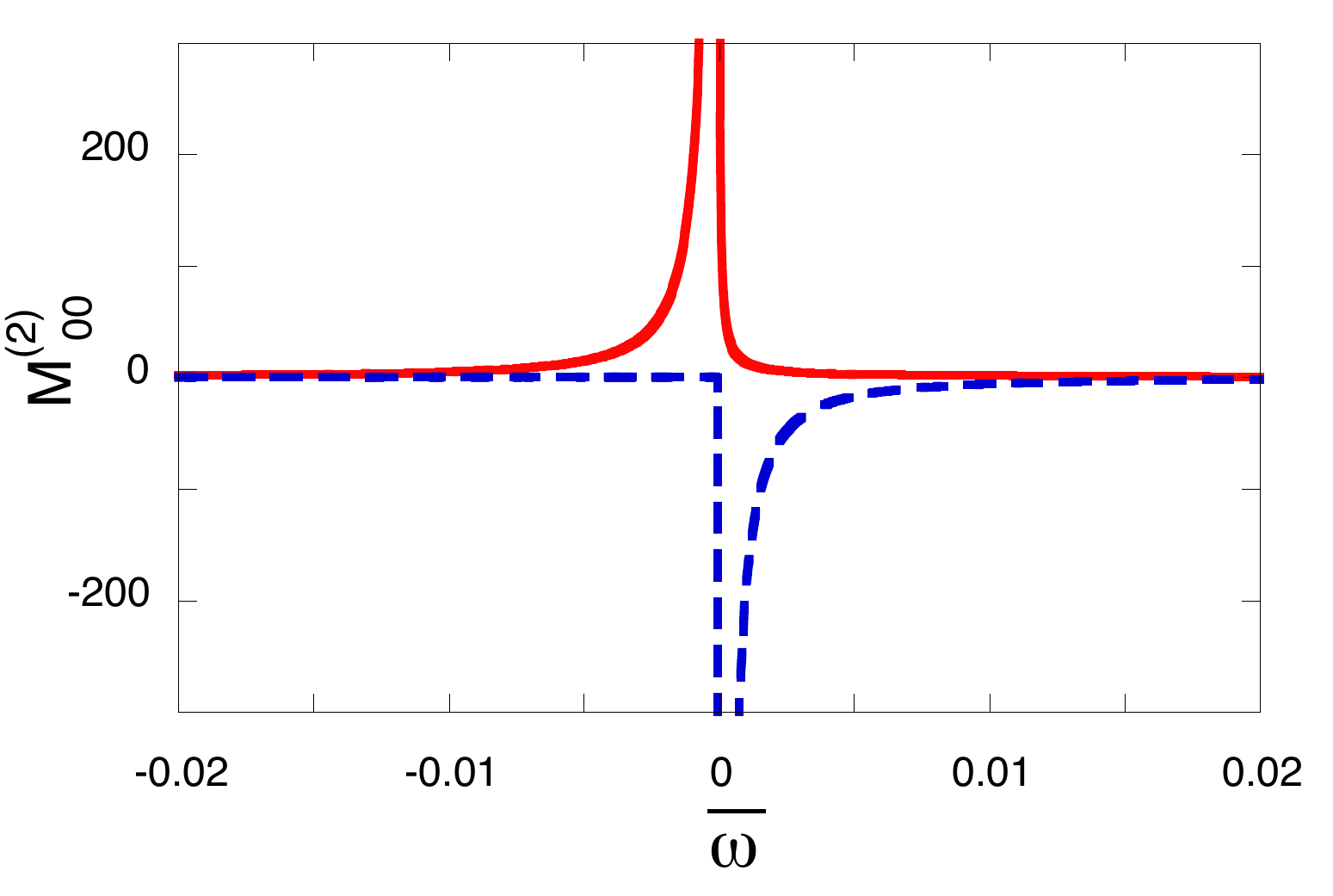}}
\caption{(Color online) Real (solid) and imaginary (dashed) parts of the correction term $\overline{M^{(2)}_{00}}=\mu a {M^{(2)}_{00}}$ vs. $\overline{\om}=\mu a^2 \om$.}
\label{fig:3}
\end{figure}
\newline
We can now formulate the adiabatic limit for a particle prepared in the ground state of a rectangular well, $m=0$, at $t\to -\infty$, provided the state "turns" before reaching the continuum threshold, $\E<0$. As in the case of the zero-range potential, the "absorbing region" is in Fig. 6 separated by a "classically forbidden region" and becomes inaccessible for a "particle" incident on the $n=0$ "potential surface" as $v\to 0$. There is, however,  a possibility to access the "absorbing potential" in Fig. 6  by hopping to a different "potential surface". But as $v\to \infty$ the hopping also becomes improbable, since the solutions on different "surfaces" become highly oscillatory, and the integrals involving $M_{0n}^{(1,2)}(\om)$ vanish. We, therefore, have the adiabatic limit
 \begin{eqnarray}\label{r9}
P^{stay}_{0n}(\E<0,v\to 0)\to \delta_{0n}.
\end{eqnarray}
This result is easily extended to other initial states, $m\ne 0$. 
%\newline
The behaviour for other values of $\E$ and $v$  requires more attention, and we will consider it next.

\section{A rectangular well: the single-Sturmian approximation}
The above discussion suggests that if the trapping potential changes sufficiently slowly, one can largely neglect scattering into 
other bound states of the well, thus leaving a few, or indeed just one, equation in (\ref{8}). For a particle arriving in the adiabatic 
ground state, $m=0$, we, therefore, write
\begin{eqnarray}\label{z1}
%\nonumber
{B_0^*}''+[v^{-2}(\E-\rho_0^*)+(S_0|\partial^2_\om S_0)^*]B_0^*=0, 
%\frac{1}{v}\left[ \frac{-\Om_0}{2}+\frac{\rho}{2}\right ]^{1/2}.
\q\q\q\q\q\q\n
%\\
%+\Om_0\sum_n M^{(0)}_{mn}B_n-\rho_mB_m(S_m(\om)|S_m(\om)),\q\q
\end{eqnarray}
where we have retained the diagonal correction term $M_{00}^{(2)}(\om)$. With no analytical 
solution available for Eq.(\ref{z1}), we have to solve it numerically.
\newline
In the dimensionless variables $\tau=4ma^2t$, $y=x/2a$, $[\tild{\E}=4\mu a^2\E$, and $\tild{v}=8\mu^{3/2}a^3v$ the SE (\ref{2}) reads 
$i\partial_{\tau}\Psi(y,\tau)=-\partial^2_y\Psi/2+[\tild{\E}-\tild{v}^2\tau^2)]\theta(y+1/2)\theta(1/2-y)\Psi$, and we must solve Eq.(\ref{z1}) for a particle of a unit mass in a well of a unit length, replacing $\om$ with $\mu a^2 \om$. The results for $P^{stay}_{00}$are shown in Fig. 8 together with the exact curves, obtained by solving numerically the original SE (\ref{2}). 

The exact results are worth a brief discussion.
For for the ground state just "touching" the continuum threshold, $\tild{\E}=0$, the $P_{00}^{stay}$ tends to the constant value in Eq.(\ref{b5}), 
 \begin{eqnarray}\label{z2}
P^{stay}_{00}(\tild{\E}=0,\tild{v}\to 0) \to 4\cos^2(2\pi/5).
\end{eqnarray}
This can be understood by scaling the variables in Eq.(\ref{2}) in a different way, so as to put to unity the particle's mass $\mu$ as well as $v$, i.e., $\tau=m^{1/5}v^{4/4}t$, i.e., $y=\mu^{3/5}v^{2/5}x$. With this we also have $W(y)=[\theta(y+\mu^{3/5}v^{2/5}a)\theta(\mu^{3/5}v^{2/5}a-y)]/2\mu^{3/5}v^{2/5}a$. As $v\to 0$,  the width of $W(y)$ tends to zero, and we recover the zero-range result (\ref{z2}), which holds universally for all values of $v$ and, in particular, for $v=1$.
\newline
For a rapidly changing rectangular trap, $\tild{v} \to \infty$, the particle always returns to the well, regardless  of whether the adiabatic state "turns" before touching the continuum, just touches it, or even disappears for a while. An yet different type of scaling can be used to explain why. Putting to unity the well's width as well as $v$, we have a particle of a mass $\tild{\mu}=4a^2v^{2/3}\mu$, and a new parameter, 
$\tild{\E}=\E/v^{2/3}$. As $v\to \infty$, we have a picture of a very heavy particle, $\tild{\mu}\to \infty$,  brought to the continuum threshold, $\tild{\E}\to 0$, and then down again. The massive particle has no chance to escape, and we have
 \begin{eqnarray}\label{z3}
P^{stay}_{00}(\tild{\E},\tild{v}\to \infty) \to 1,
\end{eqnarray}
which holds for all finite values of $\tild{\E}$.
\newline
Finally, if a bound state disappears, the particle's state is a wave packet of continuos states, which spends a duration of $2\sqrt{\E}/v$ spreading away from the region. For $v \to 0$ the time of spreading is very long, so that little is recaptured after the bound state reappears at $t=\sqrt{\E}/v$. Thus we have
 \begin{eqnarray}\label{z4}
P^{stay}_{00}(\tild{\E}>0,\tild{v}\to 0) \to 0.
\end{eqnarray}
The single-Sturmian approximation for $P^{stay}_{00}$, obtained by solving Eq.(\ref{z1}), is in good agreement with the exact result for $\tild{v} \lesssim 40$. Comparing the two curves with the total probability to stay in the well, $P^{stay}_0=\sum_{n}P^{stay}_{0n}$, shown in the inset in Fig. 8 helps identify three approximate regimes. 
%%%%%%%%%%%%%%%%%%%%%%%%%%%%%%%%%%%%%%%%%%%
\begin{figure}
	\centering
		\includegraphics[width=9cm,height=7cm]{{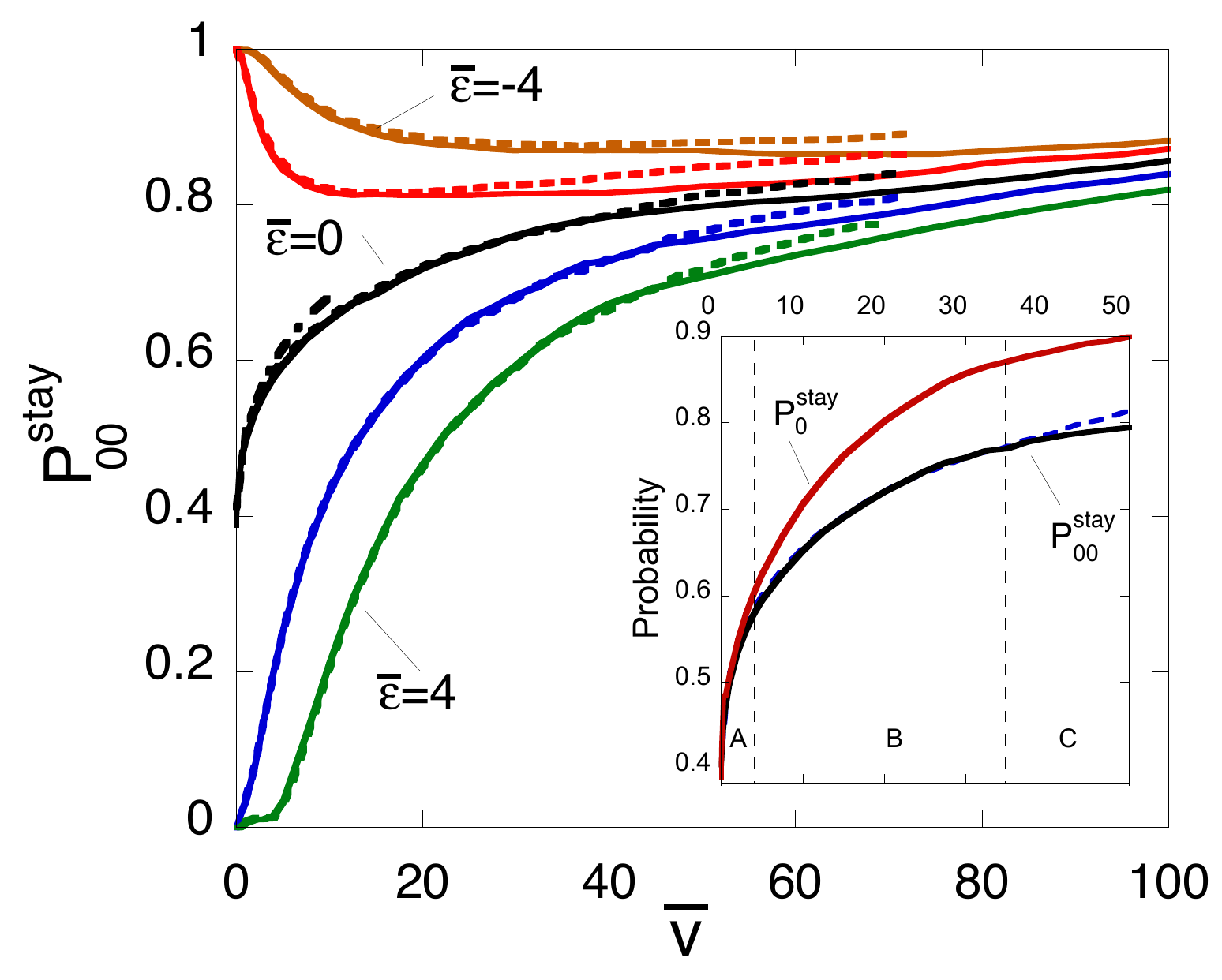}}
\caption{(Color online) Probability to remain in the ground state of a rectangular well, $P^{stay}_{00}$ vs. $\overline{v}=8\mu^{3/2} a^3v$ for $\overline{\E}=-4,-2,0,2,4$ (solid). Also shown are the single-Sturmian approximations to these probabilities obtained  with (dashed) and without (dot-dashed) the last term in Eq.(\ref{z1}). The inset shows the total probability $P_0^{stay}$ (thick solid), the exact $P_{00}^{stay}$ (solid), and the single-Sturmian approximation to $P_{00}^{stay}$ (dashed). The vertical dashed lines indicate the three regimes described in Sec. VIII. }
\label{fig:3}
\end{figure}

A) {\it Slow passage.} For $\tild{v} \lesssim 5$, we have $P^{stay}_0 \approx P^{stay}_{00}$. The loss and recapture of particles is determined by interaction of a single bound state with the continuum. There is no scattering into other bound states. Mathematically, the problem reduces to solving a single equation (\ref{z1}) (see Fig. 9a).

B) {\it Intermediate passage.} For $5 \lesssim \tild{v} \lesssim 80$, we have $P^{stay}_0 > P^{stay}_{00}$, $(P^{stay}_0 -P^{stay}_{00})/P^{stay}<<1$, with $P^{stay}_{00}$ is correctly described by Eq.(\ref{z1}). This suggests that a downward bound initial state recaptures some of the particles, and later each new bound state, which enters the deepening well, scoops some more (see Fig. 9b). This regime can be described by solving Eqs.(\ref{8}) iteratively, using the solution of (\ref{z1}) as an initial approximation.

C) {\it Rapid passage.} For $80\lesssim  \tild{v} $ we find notable discrepancies  between the single-Sturmian approximation for $P^{stay}_{00}$, and the exact result. This indicates that the loss to continuum is accompanied also by transitions between different bound states. Mathematically, this requires solution of the full "coupled channels problem" (\ref{8}) (see Fig. 9c).
We note that in the case of several spatial dimensions reduction of the original problem (\ref{2}) to that of solving a system of ordinary differential equations may be a significant simplification. 
%We will, however, limit us to the single-Sturmian approximation in this work.
\begin{figure}
	\centering
		\includegraphics[width=9cm,height=4cm]{{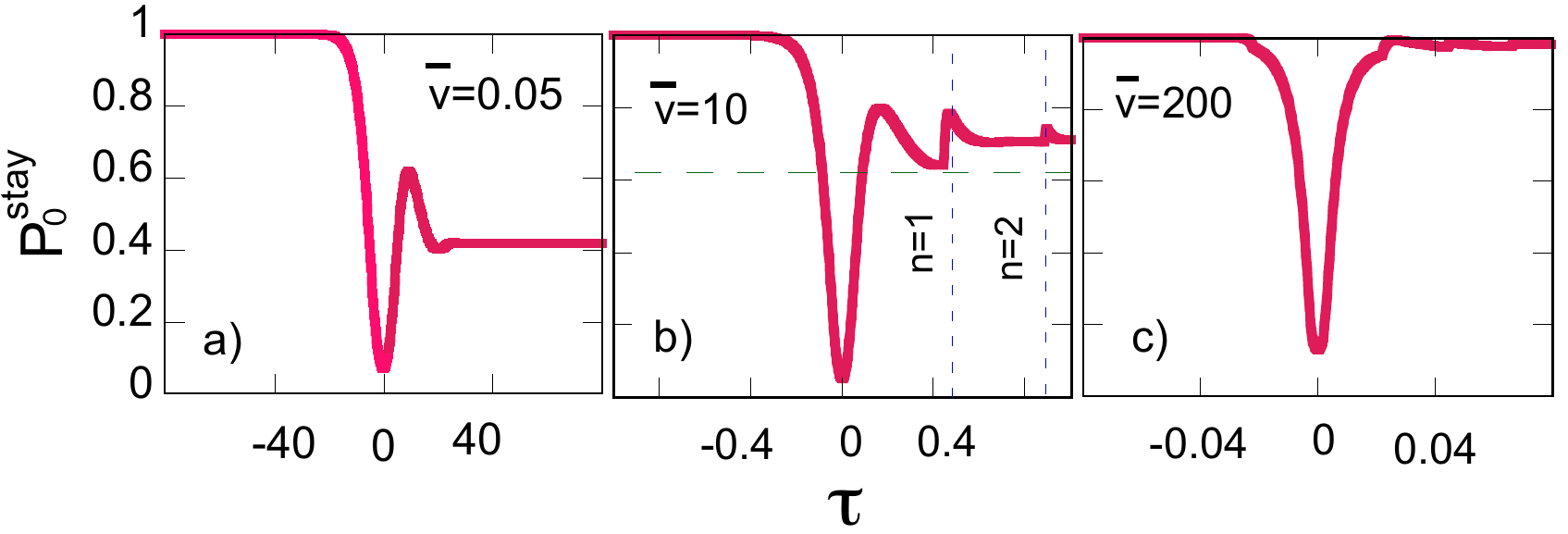}}
\caption{(Color online) Rectangular barrier: $P^{stay}_0$ vs. $\tau=4\mu a^2 t$ for $\overline{v}=0.05$; b) same as a), but for $\overline{v}=10$.
Vertical dashed lines indicate the moments when the first and the second excited states enter the well; c)same as a), but for $\overline{v}=200$.}
\label{fig:3}
\end{figure}
%%%%%%%%%%%%%%%%%%%%
\section {Universality of the "$38\%$ rule" in the $v\to 0$ limit}
We have shown that in the two cases considered above, there is no conventional adiabatic limit for a ground state just touching the continuum threshold. Rather, as $v\to 0$, the probability to remain in the state is given by Eq.(\ref{b5}), and equals approximately $38\%$. It easy to show that, for a quadratic evolution (\ref{3a}), this result holds true in one dimension for a finite-width potential of an arbitrary form. Indeed, scaling the time and coordinate so as to put to unity $v$ and the particle's mass $\mu$, while maintaining the normalisation $\int W(x)dx=1$, 
 \begin{eqnarray}\label{y1}
\tau=\mu^{1/5}v^{4/5}\tau, \q y=\mu^{2/5}v^{2/5}x, 
\end{eqnarray}
converts the SE (\ref{2}) ($V^{(0)}=0$) into
\begin{eqnarray}\label{y2}
i\partial_\tau\Psi(x,t)=-\partial^2_y\Psi/2+\tau^2 \tild{W}(y)\Psi,
\end{eqnarray}
where $\tild{W}(y)\equiv \mu^{-3/5}v^{-2/5}W(\mu^{-3/5}v^{-2/5}y)$. As $v\to 0$, we have $\tild{W}(y)\to \delta(y)$ for any choice of 
$W(x)$, so that $P^{stay}_{00}$ is given by Eq.(\ref{b5}). To illustrate this, we plotted the results for a rectangular well (\ref{r1}), and a cut-off parabolic potential 
 \begin{eqnarray}\label{y2a}
W(x)=1.5 [\theta(x+a)\theta(a-x)]x^2/a^3,
\end{eqnarray}
 in Fig. 10a.
\newline
The case where the $m$-th excited state of the well touches the continuum requires more attention.
Let the evolution of the potential be such that at $t=0$, 
%for the $m$-th bound state
 %of the well touches the continuum, 
%that is $V^{(0)}$ is chosen in such a way that, 
in the potential $-V^{(0)}W(x)$, we have $E_m=0$. Returning to Eq.(\ref{8}) we note that as $v\to 0$, the last sum in it may be neglected. Also, in this limit "absorption" of the "particle" occurs in a small vicinity of $\om=0$. Thus, if we can 
demonstrate that, for $\om\approx 0$,  $B_m(\om)$ satisfies Eq.(\ref{b1}), obtained earlier for a zero-range well, we will have also proven that $P^{stay}_{mm}(v\to 0)= 4\cos^2(2\pi/5)$, for all $m$'s and all potential shapes.
\newline
To show that this is the case, we use the standard approach, commonly used to describe potential scattering at low energies \cite{BAZBOOK}.
To this end, we will consider a potential $\rho W(x)$, whose $m$-th bound state lies just below the threshold, construct a continuum state for a small positive energy, $\om>0$, and look for the condition under which the scattering amplitude diverges, $S(E,\rho)\to \infty$. Namely, our state will have the form $\exp(\pm ikx)+S(\om,\rho)\exp(\mp ikx)$, $k=\sqrt{2\mu \om}$, for $x^<_> a$.
In the low energy limit, $ka << 1$, the wave length of the particle is large, and the well is characterised by a single parameter, the logarithmic derivative of the bound state's wave function at $x=\pm a$, which we denote  $-\kappa$, so that $\phi_m(x,\rho)\sim \exp[-\kappa(\rho)x]$. We note that $\kappa$ depends on the potential shape via $\rho$, but not on the energy $\om$, as long as $\om$ is small. Matching the log-derivatives at $x=\pm a$, and using $ka>>1$ then yields
 \begin{eqnarray}\label{y3}
S(E,\rho)=\frac{ik-\kappa(\rho)}{ik+\kappa(\rho)}, 
\end{eqnarray}
\begin{figure}
	\centering
		\includegraphics[width=9cm,height=7cm]{{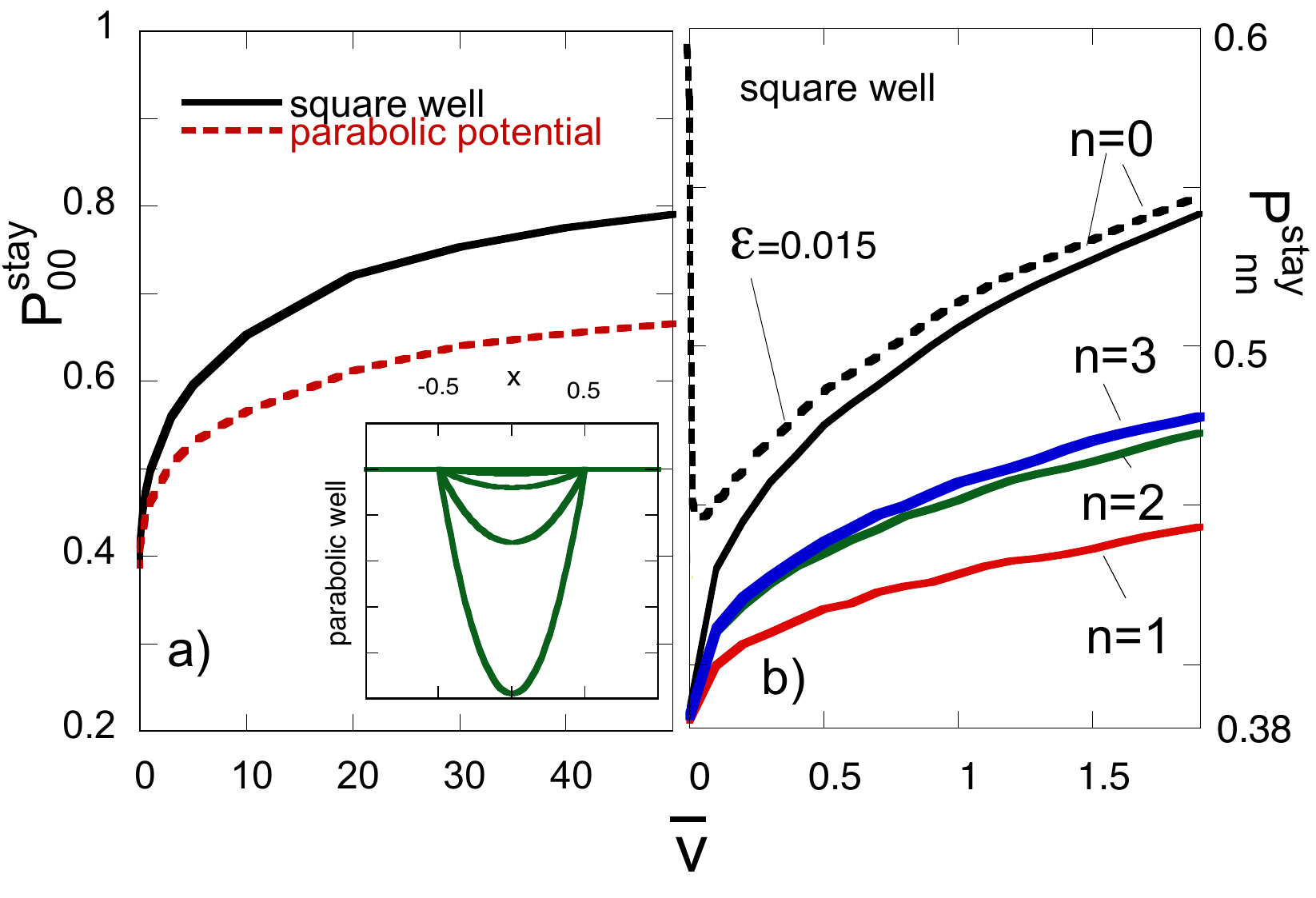}}
\caption{(Color online) a) the probabilities to remain in the ground state of the square (dashed) and parabolic (solid) wells, $P^{stay}_{00}$ for $\E=0$, vs. $\overline{v}=8\mu^{3/2} a^3v$
The inset shows different shapes of the parabolic well (\ref{y2a});
 b) same as a), but for the first $4$ states of a rectangular well "touching the continuum threshold". A dashed line shows $P^{stay}_{00}$ for $\E=-0.015$.}
\label{fig:3}
\end{figure}
which diverges whenever $ik+\kappa(\rho)=0$. The condition is usually used to obtain the pole in the complex $\om$-plane, given a real value of $\rho$ \cite{BAZBOOK}. We, on the other hand, require the value of $\rho$, given a real value of $\om$, and need to make an additional assumption about how $\kappa$ depends on $\rho$. The scattering length, defined as $\mathcal{L}=-1/\kappa$, is known to remain real, diverge, and change its sign as the shallow bound state moves toward the continuum threshold, and eventually becomes a virtual state \cite{BAZBOOK}. Thus, we will assume $\kappa$ to be a linear function of $\rho$, 
  \begin{eqnarray}\label{y4}
\kappa(\rho) \approx C(\rho-\rho^0), 
\end{eqnarray}
where $C>0$ is a real constant, and $E_m(\rho^0)=0$. Solving the pole condition $ik+\kappa(\rho)=0$ for $\rho$, we have 
  \begin{eqnarray}\label{y5}
\rho_m(\om)=i\sqrt{2\mu \om}/C_m+\rho^0_m,
\end{eqnarray}
where we recalled that our derivation is for a particle prepared in the $m$-th state of the deep well, and added the index $m$, where required. Inserting (\ref{y5}) into Eqs.(\ref{8}), neglecting all but one of them, and noting that $V^{(0)}=\rho^0_m$, for $\om \approx 0$ we have
  \begin{eqnarray}\label{y6}
{B_m}''+c\sqrt{\om}B_m=0, \q c=-i\sqrt{2\mu}/C_mv^2
\end{eqnarray}
The similarity between Eqs.(\ref{b1}) and (\ref{y6}) in the region of interest allow us to conclude that for any state $\phi_m$, $m=0,1,2..$, "touching the continuum threshold"
  \begin{eqnarray}\label{y6}
lim_{v\to 0} P^{stay}_{mn}=4\cos^2(2\pi/5)\delta_{mn}\approx =0.3819\times\delta_{mn}.\q
\end{eqnarray}
This general result is valid for any potential, provided the scattering length $\mathcal{L}$ has a simple pole when the bound state 
$\phi_m$ joins the continuum, $\mathcal{L}(\rho)\sim1/(\rho-\rho^0)$, where $E_m(\rho^0)=0$. This condition is fulfilled, for example, for  a rectangular well  (\ref{r1}), with the results for various excited states, obtained by integrating Eq.(\ref{z1}), shown in Fig. 10b. 
% \begin{eqnarray}\label{r6}
%S_n(x,\om)\sim \exp[\pm ik(\om)x], \q x>(<)\pm a.
%\end{eqnarray} 
\section{Summary and conclusions.}
In summary, we have analysed, in one dimension, the evolution of a particle prepared  in a bound state of a trapping potential, whose magnitude has a simple maximum at $t=0$, as described by Eq. (\ref{3a}). There are three possible scenarios for the  state, which first approaches the continuum threshold, and then moves away from it. It may (i) turn before reaching the continuum threshold, (ii) just touch it once, or (iii) cross the threshold and temporarily disappear. Whether the particle remains in the trap, or is lost to the continuum,  depends on how fast is the variation of the trapping potential. 
 
In the {\it slow passage} limit, the particle always remains in its initial ($m$th) state, provided the state "turns" before reaching the threshold, in accordance with the Adiabatic Theorem.
 If the state touches the threshold, the probability to remain in it
$P_{mm}^{stay}$ 
 is approximately $38\%$.
This result holds universally for all excited states and various potentials, under a very general assumption about the behaviour of the scattering length (\ref{y4}),  and replaces the conventional adiabatic limit. 
If the bound state disappears for while, a particle ejected into the continuum has sufficient time to move away from the potential.
Thus, there is $100\%$ loss to the continuum, and nothing is recovered when the state reappears.

In the {\it rapid passage} limit, the outcome depends on the choice of the potential. Thus, for a zero-range well, $P_{00}^{stay}$ tends to the same $38\%$ limit, regardless of whether the bound state turns, touches the threshold, or crosses it.
This appears to be a consequence of a perfect balance between the time a bound state of a $\delta$-well spends near the threshold, and the efficiency with which the particle is ejected. On the other hand, in the case of a rectangular potential, a rapidly evolving well always retains the particle in its original state, whichever the fate of the  bound state.

The general case of a passage which is neither slow nor fast is conveniently studied in the Sturmian representation.
%Th Sturmian method reduces 
%The adiabatic limit for a state turning below the continuum threshold is reached because, in order to be absorbed, the "particle" has to cross a "classically forbidden region". Thus, with $v$ playing the role of the Planck's $\hbar$, absorption becomes improbable in the slow passage limit $v\to 0$.  
Unless the potential changes very rapidly, it is sufficient to employ only one Sturmian state, and the task of solving the time-dependent SE (\ref{2}) reduces to that of evaluating the reflection coefficient of a complex valued "potential", where absorption of a fictitious "particle" accounts for the loss  of the real particle to the continuum.
For larger values of $v$, several Sturmian  states need to be taken into account, and the picture is that of a "particle" capable to moving on several absorbing "potential surfaces". In general, one can loosely identify three different regimes.
 If the passage is sufficiently slow, 
the state ejects the particles on its way up, and then recovers some of them on its way down.
For faster variations, the original state recovers its share of the particles, while more particles are scooped by other states, which enter the well as its depth increases.
 At yet larger $v$'s, the loss to the continuum is accompanied by scattering into other bound states, and one needs to solve a full "coupled channels problem" (\ref{8}).
 
Verification of the above theory is within the capabilities of modern experimental techniques, e.g., of the laser-based methods for containing cold atoms in quasi-one-dimensional traps.
In spite of a practical difficulty of assuring that the state just touches the threshold, this result should be amenable to an experimental verification. Figure 10b shows $P_{00}^{stay}$ for a state that turns shortly before reaching the continuum, $\E=-0.015$, closely follows the $\E=0$ curve before shooting up to its adiabatic limit $P_{00}^{stay}=1$ for very small values of $v$. Thus, the condition $\E$ can be fulfilled approximately, provided $v$ is chosen to be not too small.

%\newline
Among other advantages offered by the Sturmian technique is a simple interpretation of the adiabatic condition for a state which turns before reaching the threshold. In this case, in order to be absorbed the fictitious "particle" must first cross a "classically forbidden region" in Fig. 2. With $v$ playing the role of a "Planck's constant", this becomes improbable, if the passage is slow. 
How  $P_{mm}^{stay}$ tends to the adiabatic limit as $v\to0$ can then be studied by evaluating the corresponding phase integrals. We will consider this in our future work, together with extending the analysis to several spatial dimensions, different temporal evolutions, and the case of several identical bosons trapped in the same bound state.
%%%%%%%%%%%%%%%%
\section {Acknowledgements:}
 Support of the Basque Government (Grant No. IT-472-10), and of the Ministry of Science and Innovation of Spain (Grant No. FIS2012-36673-C03-01) is gratefully acknowledged. DS is also grateful to Gleb Gribakin and Gonzalo Muga for useful discussions.

\section{Appendix A.}
For a $t>0$, the stationary phase approximation to the integral (\ref{2a}) evaluated along the contour specified in Sect. IV is given by
\begin{eqnarray}\label{ap-1}
\nonumber
I(t)\equiv \int_{\Gamma}q(\om)^{-1/2}S(\om,x) \exp[-i\om t +i\int_{\om_0}^\om q(\om') d\om'] d\om\\
\approx [2\pi i/q(\om_s) \Phi''(\om_s)]^{1/2}S(\om_s,x)\exp [i\Phi(\om_s,t)],\q\q
\end{eqnarray}
where $\Phi(\om,t)\equiv-i\om t +i\int_{\om_0}^\om q(\om') d\om' $, and $\om_s<0$ is defined by
\begin{eqnarray}\label{ap-2}
q(\om_s)=t.
\end{eqnarray}
Given the time evolution of the magnitude of the $\delta$-potential, there are three quantities, each of which can be used as an independent variable. These are the time itself, $\tau$, the well's depth $V(\tau)=\E_0 - v^2\tau^2$, and the energy of the adiabatic bound state supported by the well, $E(\tau)=-\mu V^2(\tau)/2$. It is readily seen that $q(\om)$ in Eq.(\ref{a6}) gives the time $\tau$, at which $E(\tau(\om))=\om$
\begin{eqnarray}\label{ap1}
q(\om)=\tau(\om)=[\E-i\sqrt{2\om/\mu}]^{1/2}/v
\end{eqnarray}
Let the lower limit in the integral in the exponent in Eq.(\ref{ap-1})
%$I= \int_{\om_{0}}^\om q(\om')d\om'$ 
be $\om_0=-\mu \E^2/2$ if $\E\ge 0$, and $0$ otherwise. This ensures that $q(\om)$ is always real non-negative for $\om <0$. Changing variables $\om-\to \tau(\om)$, and integrating by parts, we have
\begin{eqnarray}\label{ap2}
\int_{t(\om_0)}^{t(\om)} \tau d\om/d\tau d\tau =\tau E(\tau)|_{\tau=\tau(\om_0)}^{\tau=\tau(\om)}-\int_{\tau(\om_0)}^{\tau(\om)} E(\tau) d\tau.\q\q
\end{eqnarray} 
With $\tau(\om_s)=t$, and either $\tau(\om_0)$ or $E(\tau(\om_0))$ vanishing, we have 
\begin{eqnarray}\label{ap3}
\Phi(\om_s)=-\int_{\om_0}^t E(\tau)d\tau.
\end{eqnarray}
For the second derivative of the phase, $\Phi''(\om_s)$, and the pre-exponential factor,  we obtain 
\begin{eqnarray}\label{ap4}
\Phi''(\om_s)=q'(\om_s)=[dE(\tau)/d\tau|_{\tau=t}]^{-1}=\q\q
\n
-[2\mu v^2 t (\Om_0+v^2t^2)]^{-1}=-[2v^2t\sqrt{-2\mu E(t)}]^{-1}.
\end{eqnarray}
%where the "+" and "-" signs are for $t<0$ and $t>0$, respectively. Finally, the pre-exponential factor is given by
\begin{eqnarray}\label{ap5}
g(\om_s)\equiv S(x,\om_s)/\sqrt{q(\om_s)}= S(x,E(t))/\sqrt{t}
\end{eqnarray}
Inserting (\ref{ap3}), (\ref{ap4}) and (\ref{ap5}) into (\ref{ap-1}), and using (\ref{a1b}) yields the term which multiplies $A_+$ in Eq.(\ref{a10}).  Equation (\ref{a9}) for a $t<0$ can now be obtained as complex conjugate of Eq.(\ref{ap1}). 
%%%%%%%%%%%%%%%%%%%%%%%%%%%%%%%%%%%%%%%%%%%%%%%%%%%%%%%%%
\section{Appendix B.}
For a rectangular potential of  a unit width, $a=1/2$, and a particle of a unit mass, $\mu=1$, we have
\begin{eqnarray}\label{ab1}
(S_m(\om)|S^{}_n(\om'))=\q\q\q\q\q\q\q\q\q\q\q\q\n
\frac{F((p_m+p_n)/2)+F((p_m-p_n)/2)}{[F(p_m)F(p_n)]^{1/2}}\equiv G(p_m,p_n),
\end{eqnarray}
where $p_m(\om)=[2(\om-\rho_m(\om)]^{1/2}$, $p_n(\om')=[2(\om-\rho_n(\om')]^{1/2}$, and $F(x)\equiv\frac{\sin(x)}{x}$.
Thus, the coupling matrix elements are given by
\begin{eqnarray}\label{ab2}
M_{mn}^{(1)}= \frac{\partial G(p_m,p_n)}{\partial p_n}\frac{d p_n}{d \om'}|_{\om'=\om}
\end{eqnarray}
and 
\begin{eqnarray}\label{ab3}
M_{mn}^{(2)}= \frac{\partial G(p_m,p_n)}{\partial p_n}\frac{d^2 p_n}{d \om'^2}
+ \frac{\partial^2 G(p_m,p_n)}{\partial p_n^2}\left (\frac{d p_n}{d \om'}\right )^2
|_{\om'=\om}\q\q
\end{eqnarray}
The divergencies of $M_{mn}^{(1,2)}$ at $\om=2$ come from the derivatives of $p_n$, which has a branching singularity at
$\om'=0$. It follows from Eq.(\ref{r4}) that,  as $\om'\to 0$, $p_0\sim \om'^{1/4}$, and $p_{n\ne0}\sim \om'^{1/2}$. Therefore, for $\om\to 0$ we obtain [$M_{mm}^{(1)}\equiv 0$ since $(S_m(\om)|S^{}_n(\om))=1$]
\begin{eqnarray}\label{ab4}
 M_{m0}^{(1)}\sim \om^{-3/4}, M_{mn}^{(1)}\sim  \om^{-1/2} \q for\q m\ne 0,n.
\end{eqnarray}
Similarly, since the first term in Eq.(\ref{ab3}) vanishes for $n=m$, 
\begin{eqnarray}\label{ab5}
 M_{00}^{(2)}\sim \om^{-1.5},\q and\q M_{mm}^{(2)}\sim  \om^{-1} \q for\q m\ne 0.\q\q
\end{eqnarray}
For $m\ne n$ the first term in Eq.(\ref{ab3}) dominates, which leads to 
\begin{eqnarray}\label{ab5}
 M_{m0}^{(2)}\sim \om^{-1.75},\q and\q M_{mn}^{(2)}\sim  \om^{-1.5} \q for\q m\ne 0.\q\q
\end{eqnarray}

%%%%%%%%%%%%%%%%%%%%%%%%%%%%%%%%%%%%%%%%%%%%%%%%%%


\begin{thebibliography}{10}
\bibitem{FOOT} For example, consider $\om < 0$, and increase the (real) depth of the rectangular well. As the well gets deeper, there will be new bound states entering it from the continuum, and then moving downwards. Recording the wellÕs depth each time the energy a symmetric bound state coincides with $\om$ will give a (real) Sturmian eigenvalue $\rho_n(\om)$. Obviously, there are infinitely many such eigenvalues.
\bibitem{CONJ} Equations for $B^*(\om)$ contain an absorbing "potential", responsible for the 
loss to the continuum. In the Eqs. for $B(\om)$, the "potential" is of an emitting kind, and their 
form is less appealing.
\bibitem{FOOT1} This is by no means the only example where solving a time dependent SE can be reduced
to a stationary scattering problem. See, for example, \cite{BAZBOOK}, Ch.7.
\end{thebibliography}
 \end{document}